\documentclass[aps,prx,showpacs,amsmath,amssymb,superscriptaddress,twocolumn,longbibliography,10pt,nofootinbib]{revtex4-2}
\usepackage{amsfonts}
\usepackage{amsmath}
\usepackage{amssymb}
\usepackage{graphicx}
\usepackage{color}
\usepackage[english]{babel}
\usepackage[utf8]{inputenc}
\usepackage{dsfont}
\usepackage{braket}
\usepackage{mathtools}
\usepackage{latexsym}
\usepackage{hyperref}
\hypersetup{
    unicode=false, 
    pdftoolbar=false, 
    pdfmenubar=true, 
    pdffitwindow=false, 
    pdfstartview={}, 
    pdfsubject={}, 
    pdfcreator={}, 
    pdfproducer={}, 
    pdfnewwindow=true, 
    colorlinks=true, 
    linkcolor=black, 
    citecolor=black, 
    filecolor=black, 
    urlcolor=black 
}
\def \be{\begin{equation}}
\def \ee{\end{equation}}
\def \bea{\begin{eqnarray}}
\def \eea{\end{eqnarray}}

\renewcommand{\phi}{\alpha} 
\DeclareMathOperator{\tr}{tr}
\begin{document}
  \title{Eigenstate thermalization in thermal first-order phase transitions}  
  \author{Maksym Serbyn}
  \thanks{These two authors contributed equally.}
\affiliation{Institute of Science and Technology Austria, Am Campus 1, Klosterneuburg, 3400, Austria}
\author{Alexander Avdoshkin}
\thanks{These two authors contributed equally.}
\affiliation{Department of Physics, Massachusetts Institute of Technology, Cambridge, MA 02139, USA}
\author{Oriana K. Diessel}
\affiliation{ITAMP, Harvard-Smithsonian Center for Astrophysics, Cambridge, MA 02138, USA}
\affiliation{Department of Physics, Harvard University, 17 Oxford Street Cambridge, MA 02138, USA}
\author{David A. Huse}
\affiliation{Department of Physics, Princeton University, Princeton, NJ 08544, USA}
   \begin{abstract}
The eigenstate thermalization hypothesis (ETH) posits how isolated quantum many-body systems thermalize, assuming that individual eigenstates at the same energy density have identical expectation values of local observables in the limit of large systems. While the ETH apparently holds across a wide range of interacting quantum systems, in this work we show that it requires generalization in the presence of thermal first-order phase transitions. We introduce a class of all-to-all spin models, featuring first-order thermal phase transitions that stem from two distinct mean-field solutions (two ``branches'') that exchange dominance in the many-body density of states as the energy is varied.  We argue that for energies in the vicinity of the thermal phase transition, eigenstate expectation values do not need to converge to the same thermal value.  The system has a regime with coexistence of two classes of eigenstates corresponding to the two branches with distinct expectation values at the same energy density, and another regime with Schrodinger-cat-like eigenstates that are inter-branch superpositions; these two regimes are separated by an eigenstate phase transition. We support our results by semiclassical calculations and an exact diagonalization study of a microscopic spin model, and  argue that the structure of eigenstates in the vicinity of thermal first-order phase transitions can be experimentally probed via non-equilibrium dynamics.
\end{abstract}
\maketitle
\section{Introduction}
The eigenstate thermalization hypothesis (ETH)~\cite{DeutschETH,SrednickiETH,RigolNature} provides a central framework for understanding thermalization in isolated quantum many-body systems, and has been extensively tested across a wide range of models~\cite{Polkovnikov-rev}. For a generic interacting quantum model, where the ETH is expected to hold,  quantum dynamics initiated from typical non-equilibrium states leads to relaxation to locally near-thermal states and the emergence of an effective hydrodynamic description~\cite{Nahum17,Huse18}.  From this perspective, systems in which the ETH is violated or requires modification represent candidates for qualitatively different dynamics and merit special attention. 

Recent progress in studying such systems --- where naive ETH does not apply --- includes many-body localization~\cite{Huse-rev,AbaninRMP} and quantum many-body scars~\cite{Serbyn:2021vc,Moudgalya_2022,Chandran23}, both of which feature distinct dynamical behavior. In many-body localized systems, the ETH breaks down due to the emergence of quasi-local conserved quantities~\cite{Huse13,Serbyn13-1}, leading to slow dynamics and the persistence of local memory of the initial state.  Likewise, in certain systems hosting quantum many-body scars, ETH-violating eigenstates can be linked to hidden symmetries~\cite{Turner2017,Choi2018} and atypical dynamics that features periodic revivals~\cite{Ho18}, provided the system is initialized in fine-tuned states.  These examples naturally raise the question of whether there exist other generic situations where ETH must be generalized or modified. 

\begin{figure*}[t]
    \centering
\includegraphics[width=0.99\linewidth]{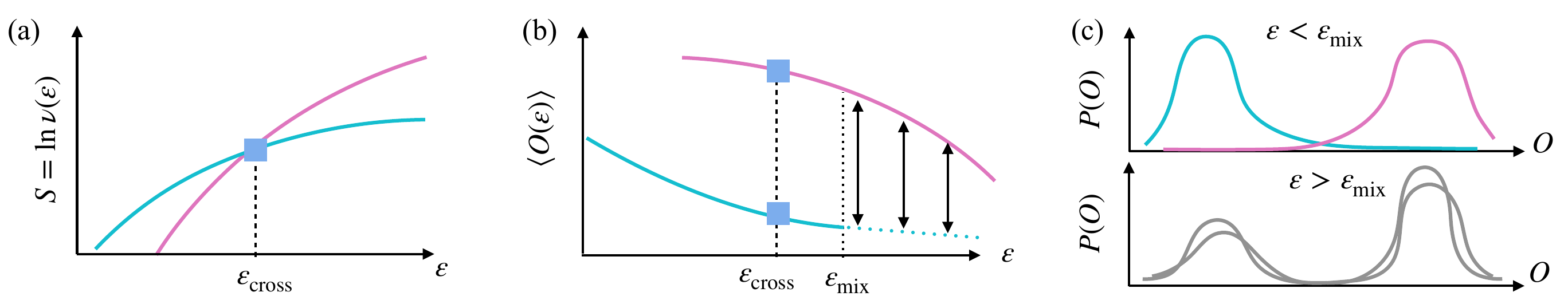}
    \caption{Schematic picture of the first-order thermal phase transition that we study.  (a) Microcanonical mean-field behavior with two distinct solutions, with one solution dominating the many-body density of states $\nu(\varepsilon)$ for energy densities $\varepsilon<\varepsilon_{\rm cross}$, and the second solution becoming dominant for larger energies. This results in a non-convex region in the entropy. 
    (b)~Expectations for the ETH, that has two different branches and absence of complete ergodicity for $\varepsilon<\varepsilon_\text{mix}$. The ergodicity is restored for $\varepsilon>\varepsilon_\text{mix}$ by a combination of classical and possibly quantum tunneling.  
    (c) The distribution of local observables in eigenstates for energy densities below $\varepsilon_\text{mix}$ has two species of eigenstates. In contrast, in the mixing regime the eigenstates have bimodal distribution of local observables.
}
    \label{Fig:cartoon}
\end{figure*}

In this work, we demonstrate that thermal phase transitions occurring in quantum systems provide a class of systems where ETH may be modified. While continuous phase transitions are more commonly studied, due to their emergent universal description, here we focus on \emph{first-order} phase transitions that are more generic~\cite{Binder87}.  In the canonical ensemble perspective, where temperature is the control parameter, a first-order thermal phase transition is  a discontinuous change in the system's energy as a function of temperature. Crucially, to study the ETH, a natural perspective is provided by the \emph{microcanonical} ensemble, that uses the energy of the system as the control parameter. When equivalence between macrocanonical and microcanonical ensembles breaks down, as it does at thermal first-order phase transitions, we understand ETH as the equivalence between quantum expectation values in eigenstates and those in the corresponding {\it micro}canonical ensemble.\footnote{We keep using the term ``ETH'' although strictly speaking the reference to temperature in eigenstate \emph{thermalization} hypothesis should be omitted when ensemble equivalence breaks down.}

In the microcanonical description the characteristic discontinuity in energy as a function of temperature is absent. Instead, the first-order phase transition is manifested as a non-convex region in the microcanonical entropy (the logarithm of the density of states) calculated as a function of energy; see Fig.~\ref{Fig:cartoon}(a) for a schematic of the behavior in the class of models that we study here. To have a model where this non-convexity in the entropy density survives in the limit of an infinite system~\cite{Binder87}, we study systems with long-range interactions.  In models with short-range 
interactions, the system removes the non-convex entropy density in the energy range of the transition by spatial phase separation. By contrast, for systems with sufficiently long-range interactions, phase separation does not occur, the entropy density can remain non-convex, and the inequivalence of the canonical and microcanonical ensembles allows the temperature as a function of the energy density to be nonmonotonic in the microcanonical ensemble~\cite{Thirring70,Gross2001}.

In this work we introduce an analytically tractable route to studying the first-order thermal phase transition in certain spin models: we will consider systems whose equilibrium thermodynamic properties follow a mean-field description and have two distinct mean-field solutions coexisting at the same energy in the energy range near the microcanonical first-order transition.  This is achieved by introducing a generalized Lipkin-Meshkov-Glick (LMG) model~\cite{LMG65} describing a collection of  spins-1/2 with all-to-all interactions.  The LMG model has symmetry under permutation of the spins which allows a mean-field description,  makes the magnitude of the total spin conserved, and reduces the dynamics to that of a single large spin. Then we perturb this model with additional weak random terms that break the permutation symmetry, allow the magnitude of the total spin to fluctuate, and make the system have true many-body dynamics, but without modifying the system's equilibrium thermodynamics.  This approach allows both analytical and numerical study of the model's ETH behavior. 

The thermal first-order phase transition is naturally obtained within the mean-field description when microcanonical entropies of two distinct mean-field solutions overtake each other with energy density, see Figure~\ref{Fig:cartoon}(a). These two branches are generally not related to each other by any discrete symmetry, and hence will have macroscopically different expectation values of local observables as shown in Fig.~\ref{Fig:cartoon}(b). The focus of the present work is to understand the fate of expectation values of local observables in individual eigenstates, therefore providing insight into ETH behavior in systems with a first-order phase transition.  Note that in general mean-field solutions with thermal first-order phase transitions need not have two branches with a crossing like this.  Instead there may be only one branch with the entropy as a function of the energy being nonconvex.  Thus what we consider here is a subset of the thermal first-order transitions that can occur in mean-field theory.

One key result of our work is the existence of the two possibilities summarized in Fig.~\ref{Fig:cartoon}(b-c). For some energies, the system may feature eigenstates that can be separated into two distinct classes according to their mean-field branch origin, the regime that we call \emph{multi-branch} ETH. In this regime, if one does not distinguish branches, the correspondence between eigenstate expectation values and the microcanonical ensemble implied by ETH breaks down. Note that such a possibility was earlier studied in the context of models with a continuous phase transition~\cite{Huse14,Srednicki15,Srednicki16,SrednickiRigol2016}, and also in quantum $p$-spin models~\cite{Scardicchio17}. Typically this regime is expected to exist at lower energy densities. 

For higher energy density we expect the quantum tunneling and thermal fluctuations that couple eigenstates on different branches to become sufficient to cause the eigenstates to be hybridized between branches. We call this regime \emph{mixed-branch} ETH, and here the equivalence between eigenstate expectation values and the microcanonical ensemble is restored. Nevertheless, we emphasize that it is qualitatively different from conventional ETH since the probability distributions of local observables for individual eigenstates are bimodal, therefore featuring anomalously large fluctuations, see illustration in Figure~\ref{Fig:cartoon}(c). Hence, in some aspects the mixed-branch ETH may be more unusual compared to the multi-branch regime. This regime was also proposed earlier~\cite{Huse14,Srednicki15,Srednicki16,SrednickiRigol2016}, however in the context of systems with $Z_2$ symmetry and a continuous phase transition, where in the multi-branch regime all eigenstates come in near-degenerate symmetry-related pairs.

The remainder of this paper is organized as follows. Section~\ref{Sec:MF} introduces the all-to-all model of spin-1/2, the resulting mean-field phase diagram, and also provides a semiclassical calculation of the tunneling that determines the fate of the ETH. This information is used in Section~\ref{Sec:ED} to  guide exact diagonalization numerics and demonstrate regimes of multi-branch and mixed-branch ETH. We conclude our work with Section~\ref{Sec:Discuss} discussing the potential implications of our results for more physical models with power-law or local interactions and outlining potential dynamical probes of the two distinct ETH regimes identified in our work. 

\section{Mean-field and semiclassical tunneling \label{Sec:MF}}

In this section we introduce a generalization of the Lipkin-Meshkov-Glick (LMG)  model that will be studied throughout this work, discuss its mean-field description and calculate the semiclassical tunneling amplitude between different mean-field solutions. Finally, we provide the mean-field phase diagrams and determine promising choices of parameters for exact diagonalization studies. 

\subsection{Generalization of the LMG model}
To explore how the ETH behaves in the presence of a thermal first-order phase transition, we consider a generalized LMG model.  The standard LMG model~\cite{LMG65,Trombettoni23} describes pairwise interacting spins with infinite-range couplings and provides one of the simplest microscopic realizations of mean-field physics in a fully quantum setting. Specifically, it is defined on a system of $N$ spin-1/2 degrees of freedom and contains single-spin terms in the form of longitudinal and transverse field, that can be expressed via collective spin operators, 
\begin{equation}\label{Eq:XZ}
X = \sum_{i=1}^N \sigma^x_i~,
\qquad 
Z = \sum_{i=1}^N \sigma^z_i~,
\end{equation}
where $\sigma_i^{x,z}$ are the standard Pauli matrices acting on the $i$-th spin. Pairwise all-to-all interactions in the standard LMG model can be expressed as a properly normalized square of the total collective spin operator, $Z^2$, leading to the Hamiltonian $H_\text{LMG} = \frac{1}{N} Z^2+ h_z Z+ h_x X$. This model, studied in numerous contexts, in particular for $h_z=0$ allows to have spontaneous symmetry breaking and study the continuous phase transition~\cite{LM06,Huse14}. However, the standard LMG model does not have a first-order {\it thermal} phase transition. The $Z_2$ symmetry pins the line of first-order phase transitions to the $h_z=0$ axis, and it cannot be crossed by increasing only the energy density or temperature. 

In order to realize a thermal first-order phase transition, we introduce a generalized 3-spin interacting LMG model, denoted as LMG-3 in what follows.  The Hamiltonian
\begin{equation}\label{Eq:LMG3}
H_\text{LMG-3} = 
\frac{1}{N^2} Z^3 + h_z Z + h_x X~,
\end{equation}
depends on two coupling parameters, $h_{z,x}$ fields, as we fixed the coupling in the 3-spin interaction term, and omitted the standard two-spin interaction.\footnote{We omit the two-spin interaction for simplicity. Including that term would not qualitatively change the result, although it breaks the chiral symmetry.}  This model strongly breaks $Z_2$ symmetry also at the level of the interaction term, and, as we show below,  has a thermal first-order phase transition for a broad region of parameters $h_x$ and $h_z$.  This Hamiltonian $H_\text{LMG-3}$ also has a chiral symmetry generated by the operator ${\cal C} = \prod_{i=1}^N \sigma^y_i$ that flips $x$ and $z$ directions of all spins. The operator $\cal C$ anticommutes with the Hamiltonian~(\ref{Eq:LMG3}), resulting in the energy spectrum being symmetric around zero energy. Due to this symmetry, we may focus only on the regime of negative energies, $E\leq 0$, as the behavior at positive energies can be obtained by application of the operator~$\cal C$. 

In addition to the chiral symmetry, this LMG-3 model also features the standard spin permutation symmetry characteristic of  models with uniform all-to-all interactions. This symmetry means the magnitude of the total spin is conserved. However, in what follows we will add a thermodynamically irrelevant random perturbation that breaks this permutation symmetry and allows the system to thermalize between different total spin sectors. Therefore, below we use the mean-field treatment to first understand the thermodynamic phase diagram of the LMG-3 model, before considering its dynamics in the presence of a weak random perturbation.
\subsection{Density matrix mean field}

\begin{figure*}[t]
    \centering
\includegraphics[width=0.999\linewidth]{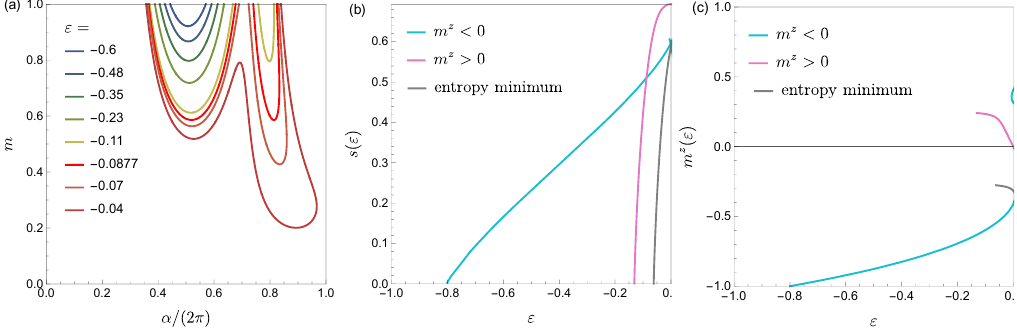}
\caption{(a) Contours of constant energy density $\varepsilon$ have a unique local minimum value of $m$ for very negative $\varepsilon$.  This minimum gives rise to the branch of mean-field solutions that we label as $m^z<0$. Upon increasing the energy density, a second local minimum of $m$ emerges. This second minimum, labeled as the $m^z>0$ branch, upon further increasing the energy attains a lower value of $m$ compared to the $m^z<0$ branch. (b) Entropy densities of the two branches as a function of $\varepsilon$ have a crossing at energy density $\varepsilon_\text{cross} \approx -0.0877$.   (c) Expectation value of $z$-magnetization per spin, $m^z$, shows that the mean-field branch appearing at lower energies has negative $z$-magnetization, while the second branch that appears upon increasing the energy has positive expectation value of~$m^z$. Grey line corresponds to the local entropy minimum (local maximum of $m$) that eventually merges with the $m^z<0$ branch at a small positive value of $\varepsilon$ (not shown). Parameters of the LMG-3 model shown here are $h_z=-0.2$ and $h_x = 0.1$. }
    \label{Fig:meanfield}
\end{figure*}

The mean-field treatment of the  LMG-3 model relies on the factorizable ansatz for the many-body spin density matrix,
\begin{equation}\label{Eq:rho}
 \rho = \otimes_i \rho_i~,
 \qquad
 \rho_i = \frac12 + \frac12 m (\cos \alpha\, \sigma^z_i+\sin \alpha \, \sigma^x_i)~,
\end{equation}
parametrized by globally uniform parameters $\alpha \in [0,2\pi]$ and $m \in [0,1]$. These parameters specify the direction and magnitude of the magnetization per spin, that reads,
\begin{equation}\label{Eq:m}
 \vec m = {\rm \tr} \rho\vec\sigma = m (\cos \alpha\, \hat z+\sin \alpha\, \hat x)~.
\end{equation}
Note that we are using only one angle $\alpha$ to specify the direction of the magnetization, since the LMG-3 model's Hamiltonian~(\ref{Eq:LMG3}) is real in the usual $\sigma^z$ basis, which means the equilibrium magnetization is in the $x$-$z$ plane. 
The entropy per spin of this mean-field state (\ref{Eq:rho}) is~\cite{Huse14}
\begin{equation}\label{Eq:s(n)}
s(m) = \frac12 [2\ln2-(1-m)\ln(1-m)-(1+m)\ln(1+m)]~,
\end{equation}
which monotonically increases as the magnetization $m$ is reduced from $1$ (maximal total spin, entropy density $s=0$) to $0$ (vanishing total spin, entropy density $s=\ln 2$).

Equipped with expressions for spin expectation values and entropy, we set up a {\it micro}canonical mean-field calculation. Specifically, at each fixed value of energy density, $\varepsilon$, given by 
\begin{equation} \label{Eq:energy-density}
    \varepsilon = \frac{H_\text{LMG-3}}{N} = 
    m^3 \cos^3 \alpha + h_z m \cos\alpha + h_x m \sin\alpha~,
\end{equation}
we seek to \emph{maximize} the entropy, or, equivalently, \emph{minimize} the magnetization $m$. The extrema of $m$, corresponding to vanishing derivative, are attained at points where $dm/d\alpha|_\varepsilon=0$~. Solving for this condition, we obtain the following expression of $m$ as a function of $\alpha$,
\begin{equation}\label{Eq:n-extreme}
 m=\sqrt\frac{  h_x \cos \alpha -h_z  \sin\alpha}{3\cos^2\alpha\sin\alpha}, 
\end{equation} 
subject to the condition of $m$ being in the physically allowed interval, $m \in [0,1]$.  Plugging this expression into Eqs.~(\ref{Eq:s(n)}) and (\ref{Eq:energy-density}) gives the values of entropy density and energy density at the extremal point as a function of $\alpha$. Generally, the entropy density is a non-monotonic function of $\alpha$ at each fixed energy, corresponding to the existence of several extremal points.  Among those extremal points, each local minimum of $m$ (local maximum of the entropy) is locally a microcanonically stable mean-field solution. 
 
In order to illustrate the solutions of the microcanonical mean-field equations described above in the regime that we will focus on, we fix parameters of the LMG-3 model to $h_z=-0.2$ and $h_x=0.1$.  With this choice of parameters, we plot the contours of constant energy density $\varepsilon$ in the plane of parameters $(\alpha,m)$ in Fig.~\ref{Fig:meanfield}(a). For strongly negative energy densities, there is a unique local minimum of $m$, with $m^z<0$, giving rise to the low-energy branch of mean-field solutions that we will call the $m^z<0$ branch.  This branch has entropy monotonically increasing with energy density, as shown in Fig.~\ref{Fig:meanfield}(b) and is characterized by a negative expectation value of $m^z$, see Fig.~\ref{Fig:meanfield}(c). Notably, upon increasing the energy density, the constant energy contours show the appearance of another local minimum of $m$.  When it first appears, we have $m=1$ at this new minimum, hence another branch of mean-field solutions enters in Fig.~\ref{Fig:meanfield}(b), initially appearing at zero entropy density near $\varepsilon=-0.13$. This branch has positive magnetization and is labeled as $m^z>0$, see Fig.~\ref{Fig:meanfield}(c). Given the chiral symmetry mentioned above, there is also another $m^z>0$ branch that is mostly at positive energy that is equivalent under this symmetry to the $m^z<0$ branch; this branch first appears very close to $\varepsilon=0$ and is slightly visible in Fig.~\ref{Fig:meanfield}(b)-(c).  We will focus only on the first two branches that appear at negative energy, so we mostly ignore this third, higher-energy branch.

When the $m^z>0$ branch first appears, it has lower entropy than the $m^z<0$ branch.  But as the energy is increased, the entropies of these two branches in Figure~\ref{Fig:meanfield}(b) cross each other at energy density $\varepsilon_\text{cross}$. So the LMG-3 model in this regime of parameters realizes the branch crossing scenario sketched in Fig.~\ref{Fig:cartoon}(a-b). These branches are characterized by macroscopically different expectation values of local observables, such as the magnetization density. The branches exchange dominance at the energy $\varepsilon_\text{cross}$, leading to discontinuities in microcanonical expectation values as function of the energy. Hence, our model realizes a strongly first-order phase transition, even in the microcanonical ensemble. Note that the all-to-all nature of the interactions does not allow the spatial phase separation that would occur in this model in microcanonical states if the interactions were instead only short-range.

\subsection{Semiclassical tunneling}
\begin{figure}[t]
    \centering
   \includegraphics[width=0.99\linewidth]{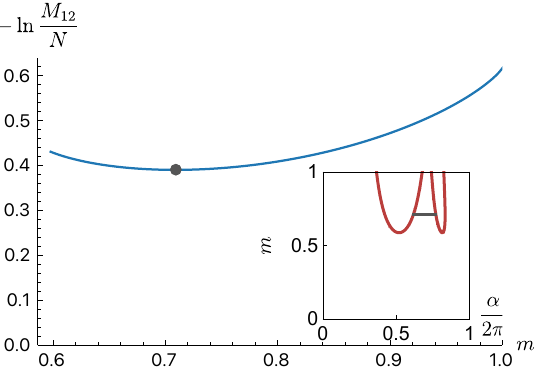}
    \caption{Tunneling at the energy density $\varepsilon_\text{cross} = -0.0877$ and $m_\text{cross}=0.586$ corresponding to the branch crossing for $h_z=-0.2$, $h_x=0.1$. The maximal matrix element is attained at a value of $m_\text{t} \approx 0.709 >m_\text{cross}$ shown by the gray dot. Inset shows the constant energy $(\varepsilon=\varepsilon_\text{cross})$ contour in the $\alpha,m$ plane, with the dark gray line indicating the value of $m_\text{t}$ and range of angles where optimal tunneling occurs.  
    } 
    \label{Fig:tunnel}
\end{figure}

Next, we consider the stability of these two coexisting mean-field solutions in the LMG-3 model with respect to tunneling processes and fluctuations.  In the LMG-3 model with exact permutation symmetry, the total spin is conserved, so does not fluctuate.  Thus now we add a weak additional term to the Hamiltonian to break this symmetry, but weak enough to not affect the thermodynamics, as discussed in more detail below.  This now allows fluctuations in $m$ within each eigenstate.  For simplicity, we mostly focus on the value of the energy density $\varepsilon_\text{cross}$, where the two mean-field solutions have the same $m=m_\text{cross}$ and thus the same entropy, see inset of Fig.~\ref{Fig:tunnel} for an illustration of this constant energy line in the plane of $(\alpha,m)$ parameters. These two extremal points may be coupled via two processes: quantum tunneling at fixed $m$ through intermediate states at different energies, and thermal fluctuations of $m$ and $\alpha$ at fixed energy.  The former process occurs within the LMG-3 model with full permutation symmetry, while the latter only occurs due to the weak breaking of that symmetry.

During quantum tunneling in this all-to-all model, the total energy of the system goes extensively ``off-shell'', so the matrix element of such processes is suppressed exponentially with the number of spins, $N$. Alternatively, the system can use thermal fluctuations to decrease the entropy (increase $m$) and go part way up or even fully over the entropy barrier staying at the same energy, a process also featuring a probability exponentially suppressed with $N$ due to the extensively lower entropy during this process.  Notably, going ``over'' the entropy barrier is possible only in cases where the two entropy maxima are separated by a local entropy minimum that is within the physically allowed $m\leq 1$ [see the gray line in Fig.~\ref{Fig:meanfield}(b)-(c)]; for the case shown in the inset of Fig.~\ref{Fig:tunnel} this process is not possible.  Generally, we expect that a certain \emph{optimal} combination, characterized by the largest net transition rate, of thermal fluctuations to a larger $m$ and quantum tunneling at that $m$ (or no tunneling if the optimum is to go over the entropy barrier),  will be the dominant process coupling the two branches in the limit of a large system. In order to find this optimum, we assume eigenstate thermalization within each branch, use the entropy to obtain the probability of the system being thermally excited to $m$, and use a semiclassical calculation of the quantum tunneling matrix element at $m$. 

A Semiclassical calculation of the tunneling matrix element, outlined in Appendix~\ref{App:SC}, gives the exponent of the quantum tunneling matrix element between the two branches at $m$, $M_{12}^{(\text{q})} \propto e^{-N A_\text{q}(m)}$, where we ignore the multiplicative factor in front of the exponential that contains polynomial dependence on $N$. The exponential suppression of the matrix element is calculated up to the leading order in $N$ via the integral over the instanton trajectory connecting the two branches at $m$. The resulting expression contains an integral over variable $z$, defined as a rescaled $z$-projection of the total spin,
\begin{multline}
\label{Eq:action-main}
A_\text{q}(m) = \frac{m}{2} \int_{z_1}^{z_2} \frac{dz}{ \sinh \varphi} \, \sqrt{\frac{1-z}{1+z}}
\\
\times \left[ \frac{6m^2z^2+h_z}{h_x} - \frac{z \cosh\varphi}{\sqrt{1-z^2}}\right],
\end{multline}
where the angle $\varphi$ is defined via the relation 
$\cosh \varphi = ({\varepsilon_\text{cross}-m^3z^3-h_z mz})/({mh_x\sqrt{1-z^2}})$, with $\varepsilon_\text{cross}$ being the energy density and $m \in [m_\text{cross},1]$. The integral is calculated between the closest points $z_{1,2}$ on the two branches, where $\cosh \varphi=1$, see Appendix~\ref{App:SC}.  

Assuming that the full matrix element for a transition between branches is the probability of thermal excitation to $m$ times the matrix element for tunneling at $m$~\cite{Huse14},  we obtain the following expression for the full matrix element,
\begin{equation}\label{Eq:Mtot}
M_{12}(m) \propto \exp \big({-N A_\text{q}(m)- N [s(m_\text{cross})-s(m)]}\big),
\end{equation}
that depends on the magnitude of the spin, $m\geq m_\text{cross}$, where the quantum tunneling process happens. The term in the exponent in Eq.~(\ref{Eq:Mtot}) has to be minimized over the values of $m$ in the physically allowed interval, $m \in [m_\text{cross},1]$.  An example of such minimization is illustrated in Fig.~\ref{Fig:tunnel}, where the gray dot indicates the optimal value of $m$, denoted in what follows as $m_\text{t}$, that generally is inside the allowed interval, $m_\text{cross}<m_t<1$. 

The resulting matrix element calculated at $m=m_\text{t}$ has to be compared to the many-body level spacing $\delta$ that is equal between branches at the crossing point, 
\begin{equation}
\delta \propto e^{-N s(m_\text{cross})},
\end{equation}
and that similarly features exponential suppression with the number of spins, $N$. The competition between level spacing and matrix element, both exponentially suppressed with $N$, leads to two distinct regimes separated by an eigenstate phase transition, as we discuss in the following. 

\subsection{Multi- and mixed-branch ETH regimes}

In the case when the optimal matrix element is much smaller than the level spacing, $M_{12}\ll \delta$, or equivalently $-(\ln {M_{12}(m_\text{t}))}/{N} > -(\ln {\delta})/{N}$, leading to condition:
\begin{equation}\label{Eq:localized}
 A_\text{q}(m_\text{t}) > s(m_\text{t})~,
\end{equation}
the tunneling process is unable to effectively couple the eigenstates from one branch to the other. Moreover the ratio between matrix element and level spacing is exponentially suppressed with the number of spins, $N$. In this \emph{multi-branch} regime, each branch is dynamically stable with respect to quantum tunneling and thermal fluctuation processes.  The eigenstates are each localized on one of the branches, so they can reliably be classified according to the branch they belong to, and any inter-branch hybridization is suppressed upon increasing the system size, $N$.  Since we are considering eigenstates at the energy where the two branches have the same entropy, this localization of the eigenstates to a single branch is a weak violation of the ETH.  We have assumed that the eigenstates do obey ETH within each branch, but in this regime they do not obey ETH once both branches are considered.

In the opposite,  \emph{mixed-branch} regime, $ -(\ln M_{12}(m_\text{t}))/N < -(\ln {\delta})/{N}$, the ratio between matrix element and level spacing is exponentially increasing with $N$. This leads to an effective mixing of the two macroscopically distinct mean-field branches. As a result, the eigenstates in this mixed-branch regime are delocalized between the branches.  Due to the macroscopically distinct character of the two mean-field branches, the eigenstates in this regime are Schroedinger-cat-like, featuring anomalous bimodal distributions of local observables. In this regime, inter-branch thermal equilibration does happen in the limit of long times, but at a rate that decreases exponentially with increasing $N$, due to the entropy barrier that has to be either surmounted or tunneled through.

Until now we have restricted our attention to the point of the crossing between two branches, where the branches have equal entropy.  Away from the crossing, the entropies differ extensively between branches, so the many-body level spacings differ, $\delta_1\gg\delta_2$, where we have labeled the lower-entropy branch 1.  The multi-branch regime, with eigenstates localized to both branches is then $\delta_1 \gg\delta_2 \gg M_{12}$. The mixed-branch regime, with full thermalization between branches is $M_{12}\gg\delta_1\gg\delta_2$.  There is also an intermediate regime, $\delta_1\gg M_{12}\gg\delta_2$, where no eigenstates are localized to the lower-entropy branch, but the eigenstates are not fully thermalized between branches because the eigenstates vary in how close they are to resonant with a state within the lower-entropy branch~\cite{Morningstar23}. There are then two eigenstate phase transitions, but the transition out of the multi-branch regime (where $\delta_1 \gg M_{12}\sim  \delta_2$) is more apparent, since this where eigenstates localized to the lower-entropy branch disappear and inter-branch relaxation onsets.

The multi-branch regime occurs when the tunneling matrix element $M_{12}$ is small, so one way to leave this regime and go through the eigenstate phase transition is to change model parameters to increase the quantum fluctuations.  In the LMG-3 model, this can be done by increasing the transverse field $h_x$ in the parameter range we study numerically below.  One can also go through the transition by changing the energy and thus the entropies of the two branches.  In the LMG-3 model, as the entropies are increased, the entropy barrier also decreases, as can be seen in Fig. 2a.  Thus by increasing the entropies and thus decreasing the $\delta$'s, we also increase $M_{12}$, so both effects go in the direction of moving across the eigenstate phase transition.

The different structure of eigenstates in the multi- vs. mixed-branch regimes can be probed via dynamics. In the multi-branch regime, initial states with local observables that agree with a particular branch will retain memory of their branch to infinite time.  Specifically, in the LMG-3 model, the $z$ magnetization initialized to be negative may never thermally equilibrate, despite not corresponding to a globally conserved quantity.  In contrast, in the mixed-branch regime, expectation values of observables will relax to their microcanonical expectation value over all eigenstates at the corresponding energy density. There, even if the magnetization is initialized to be negative, it may melt to near-zero or even positive values, depending on relative entropy of two mean-field branches. As we discussed above, these two distinct ETH regimes are separated by an eigenstate phase transition. We will provide numerical evidence for these two regimes and thus the eigenstate phase transition in Sec.~\ref{Sec:ED} below.

\begin{figure}[t]
    \centering
      \includegraphics[width=0.999\linewidth]{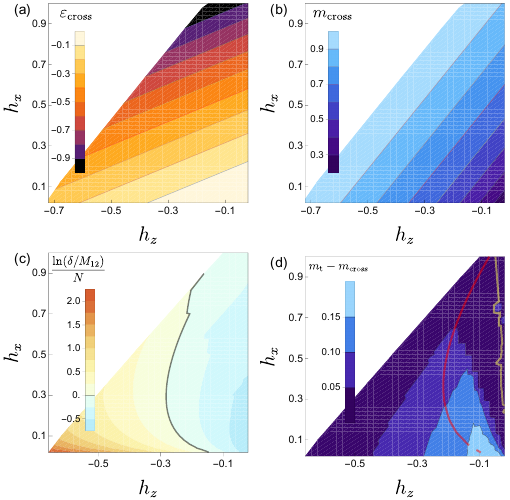}
    \caption{Phase diagrams  (a) value of energy  density $\varepsilon_\text{cross}$ and (b) spin size $m_\text{cross}$ at the branch crossing. White regions correspond to parameters where the crossing does not occur. (c) Positive/negative values of the logarithm of the ratio between level spacing and matrix element, correspond to localized and hybridizing phases, respectively. (d) Difference between optimal $m_\text{t}$ where tunneling happens and $m_\text{cross}$ at the crossing shows that the role of within-branch fluctuations is larger at small values of~$h_{x,z}$. The red line shows contour where local spin maximum at $\varepsilon_\text{cross}$ is nearly outside of the physically allowed range ($m_\text{max}=0.98$), to the left of red line only combination of quantum and classical tunneling is allowed. To the right of the yellow line tunneling becomes fully classical. }
    \label{Fig:phase-diag}
\end{figure}
\subsection{Phase diagrams \label{Sec:MF-phase}}

Finally, we combine our results to systematically study the phase diagram of the LMG-3 model as a function of the couplings $h_{z,x}$.  Our aim is to identify the most promising couplings for numerical finite-size studies of multi-branch and mixed-branch ETH regimes separated by an eigenstate phase transition in the LMG-3 model.  To this end, we perform a systematic mean-field calculation for a dense grid of points in the space of parameters $h_{z,x}$, with $h_z\in [-0.8,-0.02]$ and $h_x \in [0.02,1]$ and grid spacing $\Delta h_{x,z} = 0.02$. For simplicity, we focus on the energy density corresponding to the branch crossing point, $\varepsilon_\text{cross}$. First, in Fig.~\ref{Fig:phase-diag}(a)-(b) we show the energy density $\varepsilon_\text{cross}$ and spin size $m_\text{cross}$ as a function of the couplings. The white region in these plots correspond to the regime where the crossing is absent. Moreover, Fig.~\ref{Fig:phase-diag}(a)-(b) shows that the regime with $\varepsilon_\text{cross}$ close to zero and small $m_\text{cross}$, where the density of states is large and many such eigenstates therefore appear in finite-size systems, occur at small couplings $h_{z,x}$. We therefore focus our numerical studies on this region. 

\begin{figure*}[t]
    \centering
\includegraphics[width=0.999\linewidth]{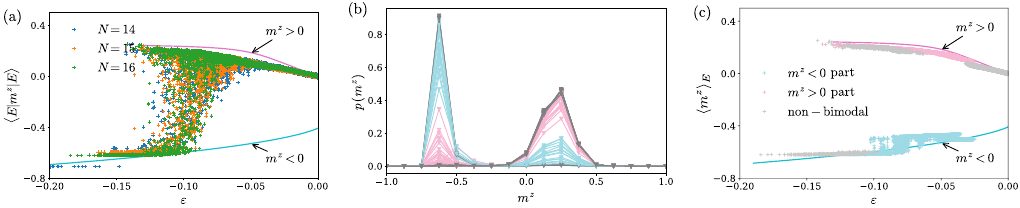}
    \caption{(a) Expectation value of $z$-magnetization for individual eigenstates for different $N=14, 15, 16$.  The mean-field branches are also shown. (b) Distribution of the magnetization for $N=16$ in the 80 eigenstates closest to energy density $\varepsilon =-0.1$ shows the presence of non-bimodal distributions, as well as bimodal distributions with peaks corresponding to the two branches. (c) After separating the bimodal eigenstates each into two parts, as described in the main text, the expectation values of the $z$-magnetization in each part, denoted as $\langle m^z\rangle_E$ clearly show the presence of the two branches for $N=16$.  Values of magnetic fields: $h_x=0.1$, $h_z=-0.2$. 
    }
    \label{Fig:ETH-example}
\end{figure*}

Next, we consider the relation between the optimal matrix element and the level spacing at the branch crossing point, see Fig.~\ref{Fig:phase-diag}(c).  Warm colors in this panel correspond to the region where the crossing point is in the multi-branch regime (localized), while blue colors mark the mixed-branch (delocalized) regime.  The accompanying Fig.~\ref{Fig:phase-diag}(d) shows the difference between $m_\text{t}$ and spin magnitude at the crossing. We observe that in a broad range of parameters the quantum tunneling is essential, since two local minima corresponding to different branches do not have a maximum of $m$ between them within physical range, similar to the case shown in the inset of Fig.~\ref{Fig:tunnel}.

The data in Fig.~\ref{Fig:phase-diag}(c) reveals that for a fixed value of the $h_z$ field near $h_z=-0.20$, increasing the transverse field, $h_x$, leads to enhanced coupling between branches and eventual delocalization. This expected trend is clear for small $h_x$, although at larger values of $h_x$ the data in Fig.~\ref{Fig:phase-diag}(c) suggests the possibility of re-entrant behavior. 
Provided that we would like the crossing point to be at small energy density, and yet potentially realize the transition between multi-branch and mixed-branch ETH regimes, Fig.~\ref{Fig:phase-diag} suggests the value of $h_z=-0.2$ to be near optimal, and for this fixed value of $h_z$, changing $h_x$ in the interval $[0.05,0.15]$ is expected to shift the crossing point from multi-branch to mixed-branch ETH regime. In the next section we proceed with an exact diagonalization study of the deformed LMG-3 model that confirms the presence of two distinct ETH regimes separated by an eigenstate transition. 

\section{Numerical study of ETH and dynamical transitions \label{Sec:ED}}
In this section we present an exact diagonalization study of eigenstate properties of LMG-3 models with up to 16 spins. To this end, we start with the discussion of the weak perturbation that we add to the LMG-3 model that is thermodynamically irrelevant but breaks the permutation symmetry and allows for numerical study of the system's ETH properties in a fully many-body regime of its dynamics.  After this we demonstrate numerical signatures of the multi-branch and mixed-branch ETH regimes and the expected transition between them. Finally, in Sec.~\ref{Sec:spec} we discuss the behavior of spectral functions of local observables and potential  signatures of the dynamical phase transition separating the different ETH regimes. While our numerical results are relatively well converged in the multi-branch and mixed-branch ETH regimes, convincing numerical characterization of the eigenstate phase transition appears to require system sizes beyond what we are able to access here.
\subsection{Model and parameter choice}
In order to break the permutation symmetry present in the uniform all-to-all LMG-3 Hamiltonian~(\ref{Eq:LMG3}), we perturb it with a random all-to-all 3-spin term,
\begin{equation}\label{Eq:random}
H = H_\text{LMG-3}+ H_\text{rand}, \quad
H_\text{rand} = \frac{\lambda}{N^p}\sum_{i>j>k} J_{ijk} \sigma^z_i \sigma^z_j \sigma^z_k,
\end{equation}
where the couplings $J_{ijk}$ are independently sampled from the Gaussian distribution with vanishing average and unit variance.  This random term, averaged over the full Hilbert space, scales as 
$\langle H_\text{rand}^2\rangle \sim N^{3-2p}$. 
Thus for $1<p< 3/2$ this term is sub-extensive and does not influence thermodynamic properties, but does grow in magnitude upon increasing $N$. In what follows we fix $p=5/4$ to be in the middle of this physically justified interval, and fix $\lambda=0.2$.

\subsection{Multi- and mixed-branch ETH regimes}

After fixing parameters, we run numerical full diagonalization on the Hamiltonian $H$ from Eq.~(\ref{Eq:random}) for three different system sizes $N=14$, $15$ and $16$, fixed value of $z$-magnetic field $h_z=-0.2$~(see Sec.~\ref{Sec:MF-phase}) and a range of values of $x$-magnetic field, $h_x$.  Unless explicitly specified we perform no disorder averaging, but use a single realization of disorder. Using numerically obtained eigenenergies and eigenstates, we study the fate of ETH in this model, focusing on expectation values of local magnetization, defined as 
\begin{equation}\label{Eq:magz}
m^z = \frac{1}{N}\sum_{i=1}^N\sigma^z_i,
\end{equation}
consistent with Eq.~(\ref{Eq:m}) used in the mean field treatment. 

The expectation value of $m^z$ across all eigenstates with negative energy density is shown in Fig.~\ref{Fig:ETH-example}(a) for $h_x=0.10$, where we focus on the range of energy densities $\varepsilon\in [-0.2,0]$, since most of the negative-energy eigenstates are concentrated in this region. The data for $-0.15<\varepsilon<-0.10$ reveals the presence of two branches, indicating that some of these eigenstates are localized within the branches.  However, as the energy density is increased past $\varepsilon>-0.1$, the lower branch of eigenstates with $\langle m^z\rangle<0$ starts to disappear. This indicates that eigenstates do not remain localized to this branch in this higher energy range, even though this lower branch does extend in Fig.~\ref{Fig:meanfield}(c) all the way to zero energy density. 

\begin{figure*}[t]
    \centering
\includegraphics[width=0.99\linewidth]{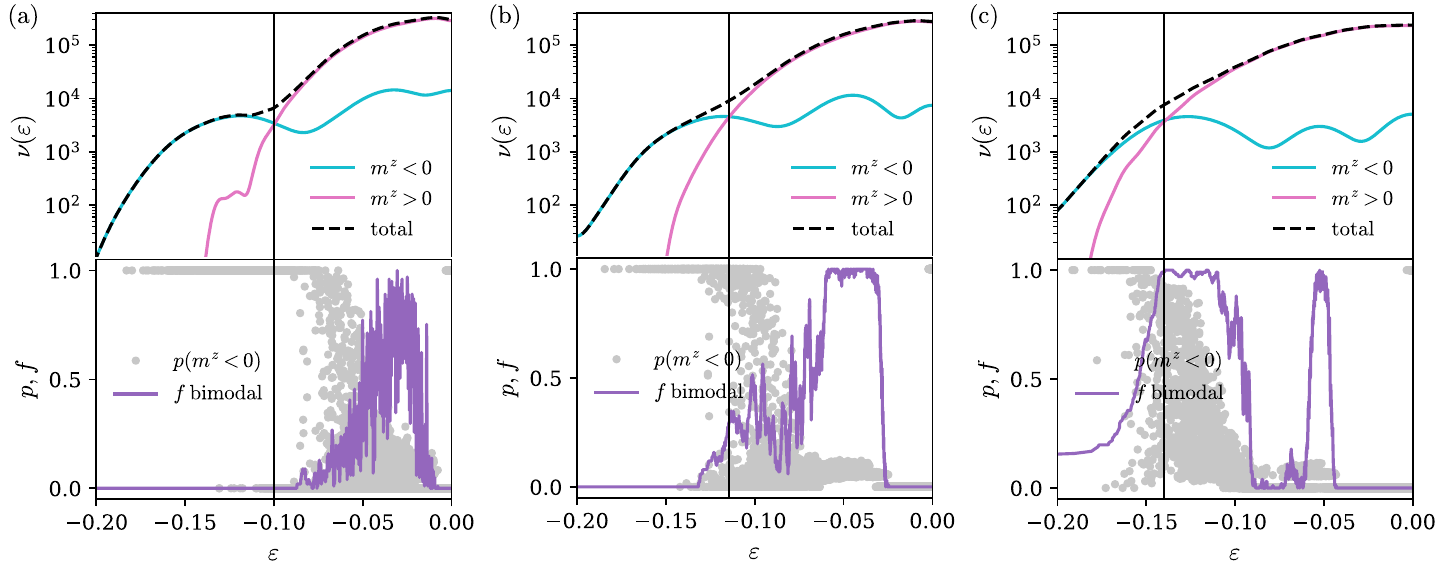}   
\caption{(a)-(c) Top panels: density of states in the two branches for three different values of transverse field, $h_x=0.05, 0.1,$ and $0.15$; the crossing moves to lower energy densities with increasing $h_x$. Bottom panels show weights of the negative magnetization peak $p$ for all eigenstates as gray dots. The magenta line gives the fraction of bimodal eigenstates $f$ (defined according to criterion $p \in [0.05,0.95]$, see App.~\ref{App:bimodal} for details). Bottom plots illustrate that with increasing $h_x$ the onset of hybridization (of bimodal eigenstates) shifts to lower energy densities. All data are for $N=16,$ $h_z=-0.2$.}
    \label{Fig:DOS-crossing}
\end{figure*}

In order to get insight into the disappearance of the negative magnetization branch, we study the distribution of $m^z$ defined in Eq.~(\ref{Eq:magz}) in individual eigenstates near energy density $\varepsilon=-0.1$. These distributions, shown in Fig.~\ref{Fig:ETH-example}(b) for 80 different eigenstates, reveal the presence of many mixed-branch eigenstates with a bimodal distribution of $m^z$, as well as other eigenstates where the distribution is non-bimodal, indicating that those eigenstates are localized within one branch. This supports the interpretation that the disappearance of the $\langle m^z\rangle<0$ branch for $\varepsilon > -0.10$ in Fig.~\ref{Fig:ETH-example}(a) is due to inter-branch delocalization of all eigenstates in this energy range, since in this range the $m^z>0$ branch has the higher entropy and thus dominates in these $m^z$ distributions.

To further explore the transition from multi-branch to mixed-branch regimes as $\varepsilon$ is increased towards zero, we demonstrate that it is possible to numerically remove the branch mixing in Fig.~\ref{Fig:ETH-example}(c).  To this end, any eigenstate that has a bimodal distribution of $m^z$ is separated into   ``$m^z<0$'' and  ``$m^z>0$'' parts by selecting the left/right peak in the distribution of $m^z$ (see Appendix~\ref{App:numerics} for details of the procedure). After proper renormalization, we plot the resulting expectation values of magnetization in Fig.~\ref{Fig:ETH-example}(c). Notably, this figure shows that such numerical removal of inter-branch mixing restores a rough agreement between mean-field and exact diagonalization results for all energy densities.  Thus, the deformed LMG-3 model for the particular choice of parameters $h_z=-0.2$ and $h_x=0.1$ as guided by the mean-field results is capable of realizing two distinct regimes of ETH, multi-branch and mixed-branch.  Below we proceed with more study of these two regimes. 

\subsection{Transition between multi- and mixed-branch regimes}
To systematically study the extent of the two different ETH regimes as a function of transverse field magnitude, $h_x$, we consider the branch-resolved density of states~(DOS). To split the total DOS into contributions from each branch, we use our classification of eigenstates into non-bimodal and bimodal (see App.~\ref{App:bimodal}). Non-bimodal eigenstates get classified according to the sign of the expectation value of $m^z$ and contribute to the DOS in only one of the branches. In contrast, bimodal eigenstates are separated into two parts, with the probability weight assigned to each branch accordingly. The resulting  per-branch DOS, as well as total DOS are shown in top panels in Fig.~\ref{Fig:DOS-crossing}(a)-(c) for three different values of $h_x$. These figures reveal that the crossing point is shifting upon increasing  $h_x$ towards more negative energy densities, consistent with the trend predicted by the mean-field results [see Fig.~\ref{Fig:phase-diag}(a)].  The systematic study of the crossing point with system size and $h_x$ is shown in Appendix~\ref{App:numerics}. 

\begin{figure}[b]
\includegraphics[width=0.99\linewidth]{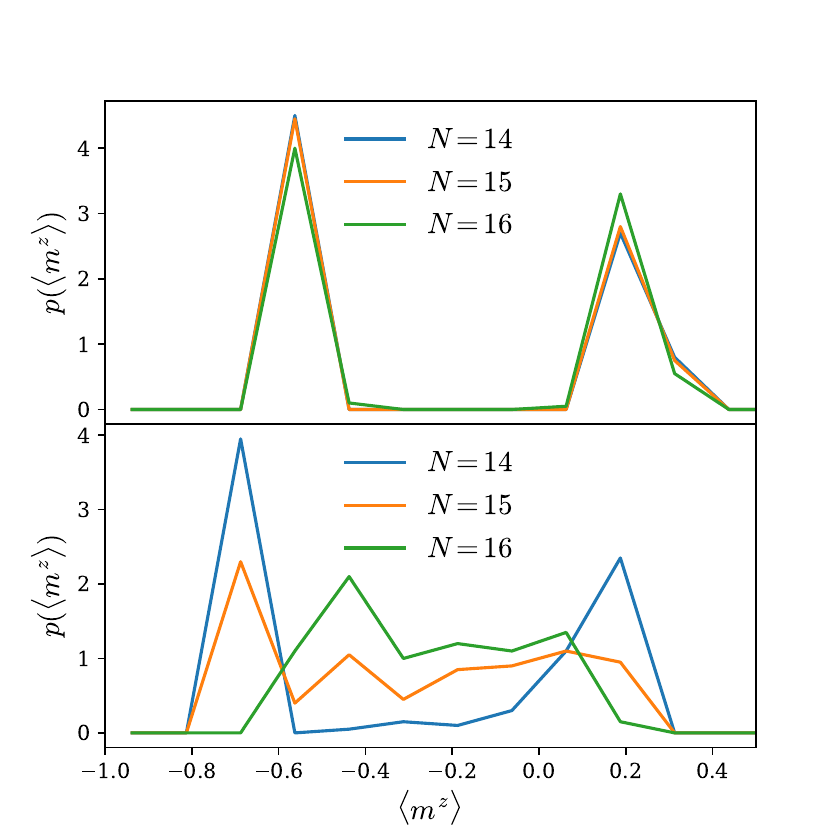}
\caption{Distribution of eigenstate magnetization expectation values at the crossing for two values of the transverse field.  For weak $h_x=0.05$ (top), this distribution is bimodal, indicating we are in the multi-branch regime,  whereas for $h_x=0.15$ (bottom) it flows towards non-bimodal with increasing system size, as expected for the mixed-branch regime.}
    \label{Fig:Z-crossing}
\end{figure}

In order to see the onset of the branch-mixing, in the bottom panels of Fig.~\ref{Fig:DOS-crossing}(a)-(c) we plot the averaged fraction of bimodal eigenstates present at each energy density, $f(\varepsilon)$ (see App.~\ref{App:numerics} for criterion and details). The onset of branch mixing is where this fraction rises from near zero to near one. We see that with increasing $h_x$ the onset of branch hybridization happens at energies above the crossing point for $h_x=0.05$ and at energies below the crossing point for $h_x=0.15$, which also qualitatively agrees with the mean-field predictions in Fig.~\ref{Fig:phase-diag}(c). 

Finally, we illustrate the behavior of the distribution of the expectation value of $\langle{E|m^z|E}\rangle$ in different eigenstates in Fig.~\ref{Fig:Z-crossing} near the energy density corresponding to the DOS crossing point. At the small value of $h_x=0.05$, this distribution stays clearly bimodal for all considered system sizes. In contrast, at the larger value of $h_x=0.15$, this distribution flows towards non-bimodal upon increasing the number of spins, $N$.  Note that at the crossing point, non-bimodal within-single-eigenstate distributions $P(m^z)$ produce the bimodal distribution seen in the top of Fig.~\ref{Fig:Z-crossing}, while bimodal within-single-eigenstate distributions produce the non-bimodal distribution seen for $N=16$ in the bottom of that figure. The presence of an eigenstate phase transition implies that the changes in this distribution and in related diagnostics should sharpen with increasing system size. However, the limited system sizes accessible to exact diagonalization can at best provide supporting evidence for such a transition and are, by themselves, inconclusive without the mean-field and semiclassical tunneling analyses presented in Sec.~\ref{Sec:MF}.

\subsection{Spectral function signatures of eigenstate phase transition \label{Sec:spec}}
To explore consequences of the branch mixing transition in dynamics, we study the behavior of off-diagonal matrix elements of the magnetization operator defined in Eq.~(\ref{Eq:magz}). We define the spectral function via rescaled off-diagonal matrix elements averaged over a narrow window of eigenstates around a target energy density, 
\begin{equation}\label{Eq:f2-def}
f^2(\omega)= \frac{N}{\delta}\left\langle |\langle E|m^z|E'\rangle|^2 \delta(E-E'-\omega)\right\rangle_{\frac{E}{N} \in [\varepsilon-\Delta,\varepsilon+\Delta]}.
\end{equation}
The rescaling is by a factor of $N/\delta$, where the inverse level spacing, $1/\delta$, is  the density of states at this energy density $\varepsilon$, and the factor $N$ appears because $m^z$ is the magnetization density. 

The spectral function defined in this way coincides with the ETH ansatz for off-diagonal matrix elements~\cite{Srednicki99,Polkovnikov-rev}. Moreover, it is related to the  Fourier transform of the real-time autocorrelation function of the magnetization density $m^z$~\cite{Srednicki99,Polkovnikov-rev}. Therefore, long-time behavior of magnetization dynamics is mapped to the low frequency behavior of $f^2(\omega)$. In particular, the  time on which dynamics of magnetization saturates, $T_\text{sat}$, sets the Thouless energy as $E_\text{Th} = 1/T_\text{sat}$ where $f^2(\omega \lesssim E_\text{Th})\sim \text{const}$. This flat region in $f^2(\omega)$ is known as the Thouless plateau in the literature~\cite{Abanin17,GG16,Abanin21}, and the corresponding energy scale is expected to be much larger than the level spacing $E_\text{Th}\gg \delta$ for systems that thermalize well. Cases where $E_\text{Th}$ is less than or near the level spacing, as is the case in strongly disordered systems~\cite{Abanin17,Abanin21,Chalker21,Roy22}  and also in some models featuring quantum many-body scars~\cite{TurnerPRB}, are typically associated with anomalous or non-ergodic dynamics. 

\begin{figure}[t]
  \centering
  \includegraphics[width=0.99\linewidth]{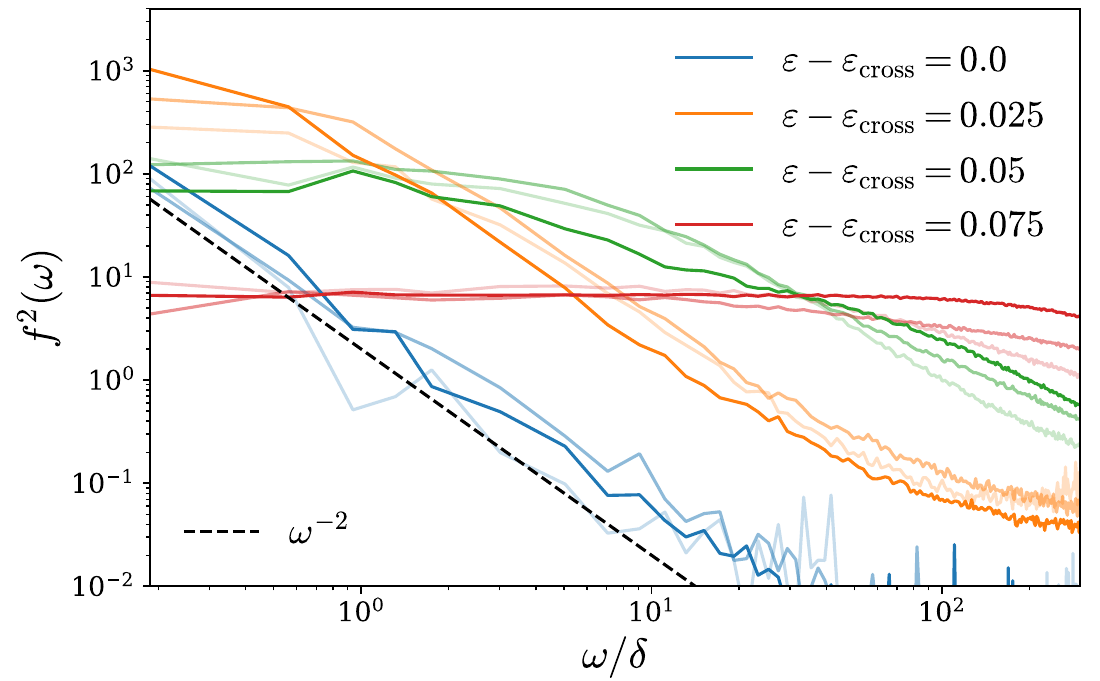}
    \caption{Rescaled spectral function of local magnetization density shows a qualitative change in behavior as a function of energy density at a fixed value of $h_x=0.1$, $h_z=-0.2$. Its value at frequency of order of the level spacing first goes up and then decreases, simultaneously showing the formation of broad Thouless plateau. Different color intensities label system sizes: 14 (fainter), 15, 16 (bolder). Energy density is plotted relative to the crossing energy density at $N=14$, and  window used for averaging is set by $\Delta=0.025$.
    }
    \label{Fig:f2w}
\end{figure}

Examples of the spectral function~(\ref{Eq:f2-def}) calculated for the perturbed LMG-3 Hamiltonian are shown in Fig.~\ref{Fig:f2w} for different energy densities. For energy densities close to zero (large values of $\varepsilon-\varepsilon_\text{cross}$), associated with mixed-branch ETH regime, the spectral function shows a broad Thouless plateau much larger than level spacing.  Thus, in the mixed-branch regime, the $z$-magnetization relaxes to its microcanonical expectation value at a rate much faster than inverse level spacing.  Upon decreasing energy density, as the system is approaching the eigenstate phase transition, we observe in Fig.~\ref{Fig:f2w} that the extent of the plateau $E_\text{Th}$  becomes of the order of the level spacing. This is consistent with the slow down in the  dynamics of magnetization relaxation at the eigenstate phase transition, $\varepsilon_\text{mix}$, where the matrix element coupling eigenstates on two different branches becomes of order the level spacing. This corresponds to the relaxation time becoming of order of inverse level spacing, and Thouless energy $E_\text{Th}\sim \delta$. 

For even smaller energy densities $\varepsilon< \varepsilon_\text{mix}$, when the system is in the multi-branch ETH regime, the magnetization does not relax to its microcanonical expectation value, but retains memory of the initial state. Such dynamics is not visible in the spectral function that is averaged over all eigenstates within a certain energy window, but can potentially be detected by distributions of off-diagonal matrix elements. The decrease in the density of states as $\varepsilon$ becomes more negative, prevents reliable averaging of the spectral function deep in the multi-branch regime, and we do not show data for $\varepsilon \lesssim  \varepsilon_\text{cross}$.

Additional support to the onset of the multi-branch ETH is given by the characteristic power-law dependence of $f^2(\omega)\sim \omega^{-2}$, shown by the dashed line in Fig.~\ref{Fig:f2w}. This dependence follows from the assumption of small matrix elements between eigenstates on different branches, $M_{12}\ll \omega\lesssim\delta_{1,2}$. The small matrix elements $M_{12}$ weakly mix the eigenstates on the two branches, with mixing amplitude estimated as $M_{12}/\omega\ll 1$ at the first order in $M_{12}$. This produces a matrix element for the magnetization that is of the order of the mixing, leading to dependence of the spectral function $f^2(\omega) \propto \omega^{-2}$, as observed in Fig.~\ref{Fig:f2w} for small values of $\varepsilon-\varepsilon_\text{cross}$. Similar power-law dependence is present also for other values of $h_x\in [0.05,0.15]$.
The power-law behavior of the spectral function was observed in the context of many-body localization~\cite{Abanin17,Abanin21} and interpreted as a signature of resonant avoided level crossings at different scales~\cite{Chalker21,Roy22}.

To conclude this section, we discussed the expectations for the dynamics of magnetization as the energy density is tuned between multi-branch and mixed-branch regimes via the eigenstate phase transition.  We supported our conclusion by numerically studying the structure of average off-diagonal matrix elements of magnetization. Crucially, the exponential density of states on each of the branches, enables the preparation of physical (short-range entangled) initial states within a narrow energy window that can be localized on individual branches. Therefore, dynamics of magnetization in different ETH regimes can be experimentally probed with quantum quenches.

\section{Discussion \label{Sec:Discuss}}
In summary, we demonstrated the existence of two distinct ETH regimes, dubbed multi-branch and mixed-branch, in certain quantum systems with a thermal first-order phase transition and all-to-all interactions.  Within the energy range of the first-order phase transition, such systems have inequivalence of the canonical and microcanonical ensembles, and eigenstate ``thermalization'' then corresponds to agreement with microcanonical distributions, and not with canonical distributions.  In the multi-branch regime, the system is characterized by incomplete ergodicity, manifesting in the existence within the same energy range of two sets of eigenstates with macroscopically distinct observable properties, that correspond to the two phases separated by the first-order transition.  In this regime, each of these two equilibrium phases is dynamically stable to infinite time even in a finite system, since almost all eigenstates are each localized primarily in one of the two phases.

In the mixed-branch regime, on the other hand, the quantum dynamics does produce slow relaxation between these macroscopically different states via a combination of quantum tunneling and thermal fluctuations, leading to a long-time restoration of full microcanonical ergodicity in large finite systems.  In this regime, the eigenstates are Schroedinger-cat-like states with the two phases that are separated by the first-order phase transition coexisting within each eigenstate with weights that match the microcanonical distribution in the limit of large systems.  The relaxation between these two ``coexisting'' phases happens on a time scale that is exponentially long in the size of the system. The multi-branch and mixed-branch regimes are separated by an eigenstate (and dynamical) phase transition, that can be probed via this exponentially slow non-equilibrium dynamics of observables. 

Our work makes a step in extending our understanding of ETH in quantum systems with thermal first-order phase transitions.  We established the mean-field language using the model with all-to-all interactions and a specific arrangement of two distinct mean-field solutions (branches). We focused on this particular regime that shows a sharp first-order phase transition as a function of energy in the microcanonical ensemble. However, qualitatively different branch arrangements can and do occur. For example, the false vacuum decay scenario that has recently attracted significant interest~\cite{Moss19,Spannowsky21,Ng21,Calabrese21,Papic25} may also be realized in all-to-all models as two branches that never cross each other.  In this scenario the eigenstates on the subdominant branch in the multi-branch regime may be interpreted as quantum many-body scarred eigenstates~\cite{Serbyn:2021vc} and they will be destroyed by the onset of inter-branch mixing.  A systematic exploration of alternative branch structures and their implications for ETH is an interesting direction for future work.

Another promising direction is to understand the existence and fate of distinct ETH regimes in models with more local interactions. Two distinct classes of such models include power-law interacting models, as well as fully local models with only short-range interactions.  Spin models with power-law interactions are implementable in trapped ion quantum simulators~\cite{Blatt12}, whereas local two-dimensional models may be realized in neutral atom or superconducting qubit quantum simulators~\cite{Devoret13}. These models allow for phase separation in the vicinity of the thermal first-order phase transition, whose effects on ETH behavior remain to be understood. Generally, we expect that unusual eigenstates and  dynamics may emerge  in the vicinity of thermal first-order phase transitions. 

Another avenue for future work is understanding the structure of off-diagonal matrix elements as well as their higher-order correlation functions~\cite{Kurchan22} related to out-of-time ordered correlation functions. To this end, it is important to find specific microscopic models realizing thermal phase transitions and amenable to numerical studies in finite-size systems. General insight provided in this direction by our work is these models should typically include more than two-spin interactions in order to strongly break the Ising $Z_2$ symmetry at the level of interactions. Alternatively, considering lattice models with three or more local Hilbert space dimensions may also be a fruitful avenue.

Finally, our findings may be also viewed in the context of metastability in quantum systems, that was recently introduced for pure quantum states in the Hilbert space~\cite{Lucas24} and also for finite temperature Gibbs states~\cite{Zhou24}. Intuitively, metastability can be defined as a class of states or density matrices for a particular model, where the energy or temperature can be only increased by the action of local operators or quantum channels. Our work suggests that metastability for mixed states can be also defined using energy instead of temperature. Within such a generalized definition, systems with a thermal first-order phase transition such as considered here have metastable states in the mixed-branch regime and stable states in the multi-branch regime, separated by the eigenstate phase transition.

To conclude, we expect that future theoretical and experimental studies of quantum models with thermal phase transitions will extend our understanding of possible ETH regimes and modifications in physical systems. These studies may also reveal generic and robust ways towards anomalously slow dynamics and unusual eigenstates at high energy densities of physical models. 

\begin{acknowledgments}
A.A. acknowledges discussions and prior collaboration on related topics with Anatoly Dymarsky. 
M.S. acknowledges Ashwin Vishwanath for introducing him to the idea of thermal first-order phase transitions in quantum systems. 
This research was supported in part by grant NSF PHY-2309135 to the Kavli Institute for Theoretical Physics (KITP) 
and by the Erwin Schrödinger International Institute for Mathematics and Physics (ESI). O.K.D acknowledges support from the NSF through a grant for ITAMP at Harvard University. 
D.A.H. was supported in part by NSF QLCI grant OMA-2120757.

\end{acknowledgments}

\appendix
\section{Details of semiclassical calculation \label{App:SC}}
In this appendix we calculate the leading terms in the semiclassical tunneling action. First we reproduce the known result for the standard LMG model, and then derive generalized expressions for the LMG-3 Hamiltonian.
\subsection{Tunneling in LMG model}
We start with reproducing the known results for the semiclassical tunneling amplitude in the conventional LMG model. The tunneling action calculation is known to have subtleties related to the formulation of coherent states path integrals for spins~\cite{Schilling1986,Grinstein92,Providencia1997,Stone03}. However since we focus on leading order terms in $N$, these subtleties do not affect the present calculation.  

We denote collective spin operators $s^{x,z}$ which differ from $X$ and $Z$ introduced in Eq.~(\ref{Eq:XZ}) by an additional factor of $1/2$. In these notations the Hamiltonian of standard LMG model may be written as $H_\text{LMG} =(g/2)(s_z^2-s_x^2)$. Translating this into Euclidean-time action we obtain, 
\begin{eqnarray}
    L &=&  i (p-s) \dot \phi - H_\text{LMG}(p,\phi), \label{Eq:L-def}
    \\
    H_\text{LMG} &=& \frac{g}{2}[(p^2-s^2)\cos^2\phi+p^2], \label{Eq:LMG}
\end{eqnarray}
where we introduced notations 
\begin{equation}\label{Eq:p-def}
p = s^z,
\end{equation}
and $\phi$ for canonically conjugate variables and remaining spin component can be expressed via them as 
\begin{equation} \label{Eq:sx-def}
    s^x = \sqrt{s^2-p^2} \cos\phi, 
    \quad
    s =  \frac{Nm}{2}.
\end{equation}

The energy in Eq.~(\ref{Eq:LMG}) has two minima confined to the $x-y$ plane of the Bloch sphere with the energy $E_0 = -s^2 g/2$. We are interested in calculating the tunneling action between these minima as 
\begin{equation}\label{Eq:A}
A = \int_{-\infty}^\infty d\tau  i (p-s) \dot \phi,
\end{equation}
using equations of motion $i\dot \phi = \frac{\partial H}{\partial p}$ and $i\dot p = -\frac{\partial H}{\partial \phi}$. These equations of motion can be integrated into explicit instanton trajectory 
 \begin{equation}
 p  = \pm \frac{i s}{\cosh (gs\sqrt{2} \tau)},
 \
 \cos^2 \phi= \frac{\cosh^2 (gs\sqrt{2} \tau)-1}{\cosh^2 (gs\sqrt{2} \tau)+1},
 \end{equation}
leading to trajectory that has $p=0$ at $\tau = \pm \infty$ and purely imaginary otherwise, and connects points $\phi=0$ and $\phi=\pi$. Explicit integration over $\tau$ gives
\begin{equation}\label{Eq:A-LMG2}
A_\text{LMG} = -s [i\pi + 2 \ln(\sqrt2+1)].
\end{equation}
The imaginary part corresponds to Berry's phase that would lead to cancellation between two different instanton trajectories for half-integer $s$. The real part of the action is correct to the leading order in $s$, and has non-trivial $O(1)$ corrections. A more direct way of calculating the action is to switch the integration variable into $\phi$ and use equations of motion with energy conservation, leading to 
\begin{equation}
    A_\text{LMG} = is \int_0^\pi d\phi\, [\sqrt{\frac{\cos^2\phi-1}{\cos^2\phi+1}}-1],
\end{equation}
that reproduces the same answer. 
\subsection{Generalization to LMG-3 model}
Here we calculate the tunneling action at the energy where two minima are degenerate in energy. In this case the energy where the tunneling happens will be given by $E_\text{min} = N\varepsilon_\text{cross}$, where $\varepsilon_\text{cross}$ is the energy density where two mean field branches have identical value of $m=m_\text{cross}$. After this, we discuss extension of the calculation to the case when tunneling happens at the higher value of $m$, corresponding to the lower entropy, but the energy and energy density are kept fixed.

Writing the generalized LMG-3 model using notations from Eqs.~(\ref{Eq:p-def})-(\ref{Eq:sx-def}), we obtain
\begin{equation}\label{Eq:LMG3-semiclassic}
    H_\text{LMG-3}
    =
     \frac{8}{N^2} p^3 +  2 h_z p + 2 h_x \sqrt{s^2-p^2}\cos\phi,
\end{equation}
where now the two distinct minima will be located on the big meridian of the Bloch sphere corresponding to $\cos \phi = -1$. In the present case we expect the instanton trajectory to have imaginary value of $\phi$ in the tunneling process, and value of $p$ to stay real. Thus, we can directly change the integration coordinate to $p$ using conservation of energy. We solve for $\cos \alpha$ using condition $H_\text{LMG-3} = N \varepsilon_\text{cross}$ with Hamiltonian from Eq.~(\ref{Eq:LMG3-semiclassic}) as:
\begin{equation} \label{Eq:angle-from-H}
    \cos \phi = \frac{ \varepsilon_\text{cross}-8 p^3/N^3-2h_z p/N}{2h_x \sqrt{(s/N)^2-(p/N)^2}}.
\end{equation}

\begin{figure}[t]
    \centering 
\includegraphics[width=0.99\linewidth]{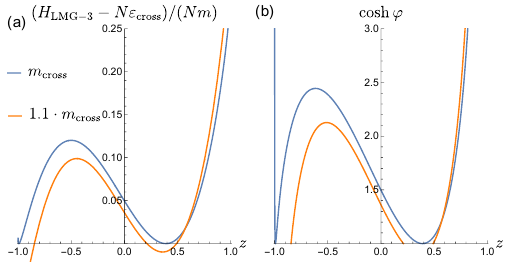}
    \caption{(a) Effective energy potential as a function of $z$ for fixed $\cos \alpha=-1$ has two minima at zero when $m=m_\text{cross}$. Tunneling happens between points $z_1 = 0.9926$ and $z_2 =0.3909$. For larger values of $m$,  for instance $m=1.1 m_\text{cross}$, tunneling happens between closest roots of effective potential that move closer together. (b) Value of $\cosh\phi$ as a function of $z$ on the instanton trajectory is consistently larger than $1$ between tunneling points, indicating that the value of original angle $\alpha$ is imaginary under the tunneling barrier.}
    \label{Fig:instanton}
\end{figure}

Under the energy barrier $|\cos \phi|>1$, hence corresponding to imaginary $\phi$. Denoting $\alpha = i \varphi$ we get:
\begin{equation}
A_\text{LMG-3} = -\int_{-\infty}^\infty d\tau\, (p-s) \dot\varphi = 
 \int_{p_1}^{p_2} dp \, (s-p) \frac{\dot\varphi}{\dot p},
\end{equation}
where we change the integration variable from time to $p$, and integration limits are given by two solutions of $\cos\phi=-1$ of Eq.~(\ref{Eq:angle-from-H}). The ratio $\dot\varphi/\dot p$ can be simplified using equations of motion, which after introducing rescaled variable 
\begin{equation}\label{Eq:z-def}
    z=\frac{p}{s} = \frac{2p}{Nm},
\end{equation}
rescaling an overall prefactor of $-N$ and substituting $m=m_\text{cross}$, gives 
\begin{multline}
    \label{Eq:action}
A_\text{q} = - \frac{A_\text{LMG3}}{N}=  \frac{m_\text{cross}}{2} \int_{z_1}^{z_2} \frac{dz}{ \sinh \varphi} \, \sqrt{\frac{1-z}{1+z}}\\
\times \left[ \frac{6m_\text{cross}^2z^2+h_z}{h_x} - \frac{z \cosh\varphi}{\sqrt{1-z^2}}\right],
\end{multline}
where $\cosh\varphi$ and $\sinh \varphi$ are functions of  $z$ given by:
\begin{equation}\label{Eq:cosh}
    \cosh \varphi = \frac{\varepsilon_\text{cross}-m_\text{cross}^3z^3-h_z  m_\text{cross} z}{m_\text{cross} h_x\sqrt{1-z^2}}.
\end{equation}
Figure~\ref{Fig:instanton} shows the example of the potential and plots the function $\cosh \alpha$ obtained from equations of motion for specific values of $h_x=0.1$ and $h_z=-0.2$ used in the main text. The numerical calculation of the integral in Eq.~(\ref{Eq:action}) gives the value  $A_\text{q} = 0.446$.

However, tunneling is possible not only from the bottom of the potential well shown in Fig.~\ref{Fig:instanton}(a). As we discuss in the main text, the system can use classical fluctuations to decrease entropy (increase value of $m$). This incurs additional suppression from classical tunneling, however decreases quantum tunneling action.  Thus, we generalize the expression for the action when tunneling process occurs at $m>m_\text{cross}$. The resulting expression (also given in the main text in Eq.~(\ref{Eq:action-main})) is identical to Eq.~(\ref{Eq:action}),  however, with value of $m_\text{cross}$ replaced by $m$ that is allowed to take values in the interval $m\in [m_\text{cross},1]$. We emphasize that we keep the value of energy density constant, as it is fixed within the microcanonical ensemble. Nevertheless, increasing value of $m>m_\text{cross}$ effectively lowers the energy barrier as shown in Fig.~\ref{Fig:instanton}(a), where for $m>m_\text{cross}$ now tunneling happens between two values of $z$ that are closest to each other. Inside this interval of values of $z$ the expression of $\cosh \varphi$ becomes larger than one, see Fig.~\ref{Fig:instanton}(b), indicating that angle $\alpha$ is imaginary on the instanton trajectory. 

Finally, we note that spin lives on the Bloch sphere, and hence the tunneling between two minima may in principle be also possible using instanton trajectory going in the other direction on the big meridian of the Bloch sphere. However, in contrast to conventional LMG model that has $Z_2$ symmetry, in the present case we have no symmetry that relates the action of two tunneling trajectories. Hence the chosen instanton trajectory is expected to give dominant contribution to tunneling.

\section{Additional numerical data}
\label{App:numerics}
In this appendix we present details of the numerical procedures used in the main text, as well as additional numerical data supporting our conclusions. 
\subsection{Determination of bimodal eigenstates \label{App:bimodal}}

Here we discuss the algorithm and criteria used to determine the bimodal eigenstates (see Fig.~\ref{Fig:ETH-example}) and also to plot their fraction (Fig.~\ref{Fig:DOS-crossing}). In order to analyze the character of eigenstates, we first calculate the distribution of magnetization $m^z$ for each individual eigenstate. Magnetization $m^z$ can take a small discrete set of $N+1$ values for $N$ spin-1/2 degrees of freedom, $m^z = -1,-1+2/N,-1+4/N \ldots,1$. The resulting distribution of magnetization is smoothed using function \texttt{gaussian\_filter1d} from \texttt{scipy.ndimage} package. Afterwards we apply peak finding function \texttt{find\_peaks} from   \texttt{scipy.signal} with parameters \texttt{prominence}$=0.01$, and  \texttt{distance}$=2$. Provided this function returns two peaks, the eigenstate is denoted as bimodal, and we also extract the location of minimum separating two peaks, used below for obtaining per-branch DOS.

The minimum separting two peaks for bimodal eigenstates is used to extract the weight in the left peak, denoted as $p$. This weight is used to plot the fraction of bimodal eigenstates, defined as those where $p \in [0.05,0.95]$, in bottom panels of Fig.~\ref{Fig:DOS-crossing} after sliding window average with window fixed to 101 eigenstates.

\subsection{Crossing point in DOS \label{App:cross}}

\begin{figure}[t]
    \centering
\includegraphics[width=0.99\linewidth]{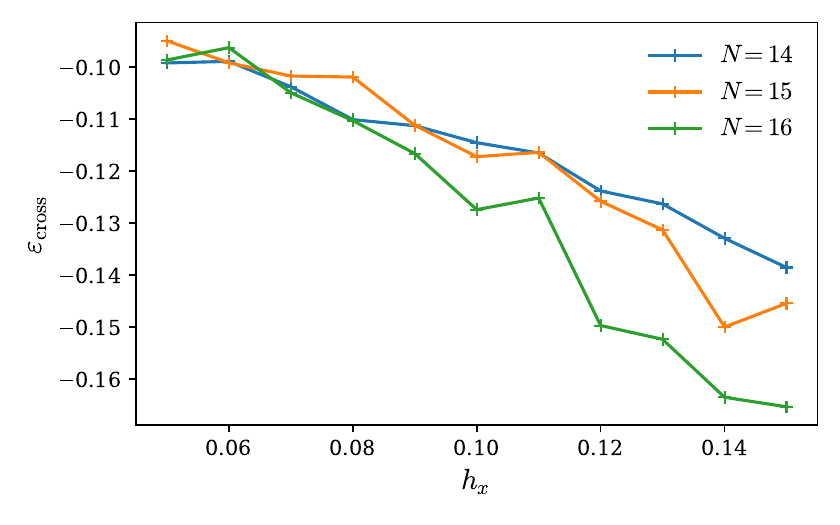}
    \caption{Energy density where two mean field branches cross each other as a function of $h_x$ for three different system sizes. }
    \label{Fig:e-cross-hx}
\end{figure}

Here we systematically study the behavior of the crossing point in DOS of two branches. The branch-resolved DOS is obtained using the algorithm for determining bimodal eigenstates described above. Eigenstates not identified as bimodal are fully assigned to one of the branches depending on the expectation value of magnetization. For the  bimodal eigenstates the distribution of $m^z$ has two clearly defined peaks separated by the minimum, see Fig.~\ref{Fig:ETH-example}. We use the location of the minimum to separate the contribution of the two peaks: the weight to the left of the minimum including the minimum point is assigned to the $m^z<0$ branch that is denoted as $p$, and the weight to the right is attributed to the $m^z>0$ branch, being equal to $1-p$ by normalization. These two weights by definition sum to one, hence, the sum of two DOS for individual branches equals to the total DOS. 

In order to determine crossing point the resulting discrete dataset of energies of individual eigenstates and their per-branch weights is smoothed using the \texttt{gaussian\_kde} from \texttt{scipy} package using bandwidth method with bandwidth set to 0.1. The crossing point of two DOS is extracted numerically from this data, see Fig.~\ref{Fig:DOS-crossing} as an example, and plotted in Fig.~\ref{Fig:e-cross-hx}. The location of the crossing point systematically shifts to lower energy densities upon increasing $h_x$, consistent with the mean field phase diagram, see Fig.~\ref{Fig:phase-diag}(a). In addition, the crossing point also shifts with $N$ indicating presence of finite size effects and displays weak fluctuations with different realizations of disorder not shown here. 

\begin{figure}
    \centering
\includegraphics[width=0.99\linewidth]{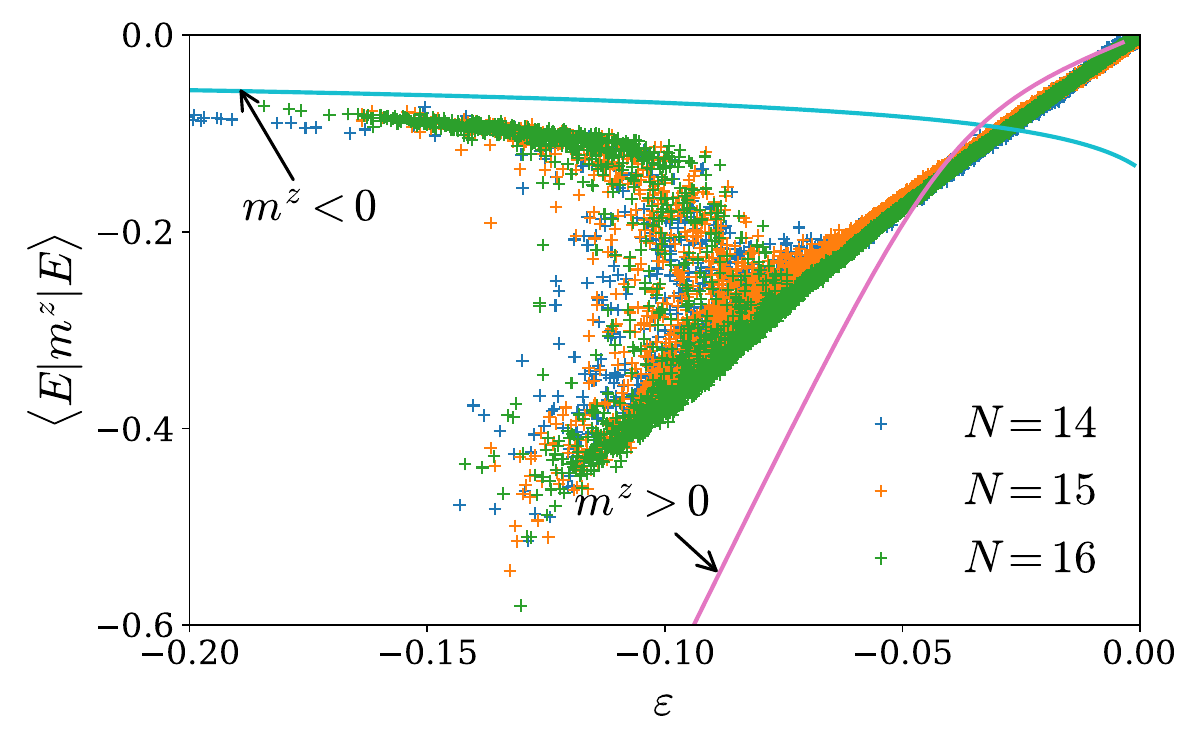}
    \caption{Expectation value of $m^z$ in eigenstates for $h_x=0.1$, $h_z=-0.2$ for three different system sizes. Lines correspond to mean field prediction, with branch labeled as $m^z<0$ featuring small negative $m^x$ values and branch with $m^z>0$ having considerably larger magnetization in $x$-direction. }
    \label{Fig:x-obs}
\end{figure}

\subsection{Diagonal matrix elements of $m^x$}

In the main text we focused on studying matrix elements of $m^z$ operator, and observed two distinct behaviors denoted as multi-branch and mixed-branch regimes. In order to illustrate that this behavior holds for other general operators, we study expectation values of $m^x$ magnetization in eigenstates. Figure~\ref{Fig:x-obs} shows these expectation values along with mean field predictions in the same interval of energy densities as Fig.~\ref{Fig:ETH-example} in the main text. While we observe a larger discrepancy between mean field predictions and expectation values in eigenstates, the overall trend is similar: at lower energy densities we see two distinct branches, that start considerably hybridizing around $\varepsilon\approx 0.1$ and for higher energy densities only the branch that dominates the DOS (labeled as $m^z>0$) survives.

\bibliography{potts}
\end{document}